\documentclass[preprint, plb, showpacs, floatfix, superscriptaddress, nofootinbib]{revtex4}
\usepackage[dvips]{graphicx}
\usepackage{epsf}
\usepackage{amsmath}
\usepackage{amssymb}
\usepackage{dcolumn}
\usepackage{bm}
\usepackage{color}
\usepackage{epsfig}
\usepackage{amsfonts}
\usepackage{enumerate}
\usepackage{setspace}
\usepackage{hhline}
\usepackage[toc,page]{appendix}

\begin{document}

\begin{flushright}
RESCEU-18/14
\end{flushright}

\title{Primordial black holes from temporally enhanced curvature perturbation}

\author{Teruaki Suyama}
 \affiliation{Research Center for the Early Universe (RESCEU), Graduate School of Science, 
The University of Tokyo, Tokyo 113-0033, Japan}
 \affiliation{Physics Division, National Center for Theoretical Sciences,
Hsinchu 300, Taiwan}

\author{Yi-Peng Wu}
 \affiliation{Research Center for the Early Universe (RESCEU), Graduate School of Science, 
The University of Tokyo, Tokyo 113-0033, Japan}
 \affiliation{Department of Physics, National Tsing Hua University,
Hsinchu 300, Taiwan}

\author{Jun'ichi Yokoyama}
 \affiliation{Research Center for the Early Universe (RESCEU), Graduate School of Science, 
The University of Tokyo, Tokyo 113-0033, Japan}
\affiliation{Kavli Institute for the Physics and Mathematics of the Universe (Kavli IPMU), 
TODAIS, WPI, The University of Tokyo, Kashiwa, Chiba 277-8568, Japan}


\begin{abstract}
Scalar field with generalized kinetic interactions metamorphoses depending on its field value,
ranging from cosmological constant to stiff matter. 
We show that such a scalar field can give rise to temporal enhancement of the curvature
perturbation in the primordial Universe,
leading to efficient production of primordial black holes while the enhancement persists.
If the inflation energy scale is high,
those mini-black holes evaporate by the Hawking radiation much before Big Bang nucleosynthesis
and the effective reheating of the Universe is achieved by the black hole evaporation.
Dominance of PBHs and the reheating by their evaporation modify the expansion history of the primordial Universe.
This results in a characteristic feature of the spectrum of primordial tensor modes
in the DECIGO frequency band, opening an interesting possibility of testing PBH reheating scenario 
by measuring the primordial tensor modes.
If the inflation energy scale is low, the PBH mass can be much larger than the solar mass.
In this case, PBH is an interesting candidate for seeds for supermassive black holes 
residing in present galaxies.
\end{abstract}
\maketitle

\section{Introduction}
Precise measurements of the cosmic microwave background (CMB) anisotropies
lead to a major progress in understanding the nature of inflaton \cite{Ade:2013uln}. 
This scenario underlies the assumption that the curvature perturbation at the epoch of last scattering 
is directly generated from the quantum fluctuations of the inflaton field. 
Ever since the curvature perturbation is generated during inflation, 
it preserves a constant value on large scales (for instance, \cite{mukhanov}).

It has been found, on the other hand, 
that a light field with insignificant energy density during inflation may affect 
the dynamics of the curvature perturbation \cite{Mollerach:1989hu,Kofman:1986wm,Yokoyama:1995ex}. 
In the curvaton model \cite{Enqvist:2001zp,Lyth:2001nq,Moroi:2001ct}, 
a representative scenario alternative to the standard one,
a canonical massive scalar field called curvaton oscillates about its potential minimum in
the radiation dominated era after curvaton mass becomes comparable to the Hubble parameter.
Such an oscillating curvaton behaves as dust and it gradually dominates the energy density 
in a radiation dominated era.
Fraction of curvaton energy density becomes maximal just before the curvaton decays 
into other relativistic particles
and the curvature perturbation is dominantly generated in this period \cite{Lyth:2002my}.
Once generated, the curvature perturbation remains constant until it reenters the Hubble horizon.
Notice that the total amplitude of the curvature perturbation is a monotonically increasing function of time. 
This fact is true even if the initial curvature perturbation generated by 
inflaton is taken into account \cite{Lyth:2005sh}.

On the other hand, it has been found that the rise of a scalar field from a subsidiary 
energy density may occur solely due to its kinetic energy.
By introducing a non-canonical kinetic interaction, $k$-essence exhibits a cosmic fluid 
with a time varying equation of state that depends on the dominant energy density of the universe \cite{Chiba:1999ka,ArmendarizPicon:2000dh,ArmendarizPicon:2000ah}. 
Attractor solutions are known to exist, which drive the scalar field to behave as a cosmological constant
while being subdominant. 
Indeed, this class of model is incorporated in the kinetic gravity 
braiding theory \cite{Deffayet:2010qz,Kimura:2010di} 
(also known as generalized Galileon \cite{Deffayet:2011gz}), 
where an evolution crossing into the phantom regime is realized without instability.
In either case, the scalar field may acquire a negative pressure so that its density increases with the cosmic expansion.
Suppose such a field exists during inflation, it can easily dominate the universe in the primordial Universe
to start the second inflation or can change into stiff matter($w=1$) 
which decays faster than radiation \cite{ArmendarizPicon:1999rj,Kobayashi:2010cm}.

In this work, we consider a spectator field, a scalar field
with negligible energy density during inflation, which has generalized kinetic interactions. 
Due to the rapid cosmic expansion, this spectator field evolves into the attractor 
that gives a negative pressure to the field. 
As a result, energy density of this field relatively increases in the post-inflationary epoch.
We explore a practical example in which the spectator field undergoes a transition from the negative pressure fluid
to stiff matter through the internal field evolution during the radiation dominated epoch. 
Given that the relative energy density of the spectator field takes a maximum during the transition, 
the curvature perturbation in this scenario exhibits a temporal enhancement, 
which is prohibited in the original curvaton scenario \cite{Lyth:2005sh}.

Although the curvature perturbation temporally generated by the spectator field is not 
observed in the CMB anisotropies, 
a temporarily large curvature perturbation leads to formations of primordial black holes (PBHs) 
at the epoch when such perturbation reenters the Hubble horizon \cite{Suyama:2011pu}. 
In the model we consider in this paper, PBH mass is small so that PBHs evaporate by the Hawking radiation
much before Big Bang nucleosynthesis.
If PBHs dominate the Universe before evaporation, radiation generated by the BH evaporation
overwhelms the original radiation of inflaton origin 
and all the matter in the present Universe comes from the PBHs.
In this sense, the effective reheating of the Universe occurs when the PBHs evaporate.
Dominance of PBHs and the reheating by their evaporation modify the expansion history of the primordial Universe.
This results in a characteristic feature of the tensor mode spectrum as 
the red-tilt ($\Omega_{\rm gw} \propto f^{-2}$)
slope in a frequency range corresponding to PBH dominated epoch and 
the (almost) flat spectrum outside that range.
Interestingly, this range naturally falls into the DECIGO band and thus our scenario can be tested by
measuring the primordial tensor modes.

The rest of the paper is organized as follows. 
In Section~\ref{model}, we define model of a spectator field we study in this paper and
solve the background dynamics both during and after inflation.
In Section~\ref{cur}, we first study perturbation of the spectator field and then
compute the curvature perturbation sourced by the field.
In Section~\ref{pbhs}, we discuss the PBH production from the temporally enhanced curvature perturbation
and its observational test.
Finally, Section~\ref{conc} is devoted to conclusion.

\section{Scalar field with generalized kinetic energy}
\label{model}
\subsection{Model}
In this section, we aim to show a model of a spectator scalar field $\phi$ that is not 
an oscillating massive scalar but gradually occupies a significant
fraction of energy density in the post-inflationary epoch. Specifically,
we consider the case in which $\phi$ has a 
noncanonical kinetic interaction $K(\phi,X)$ that vanishes as 
$X\equiv -g^{\mu\nu}\nabla_\mu\phi\nabla_\nu\phi/2\rightarrow 0$. 
This is to say that $\phi$ is exactly massless without any potential,
in contrast to the previously mentioned curvaton field.

Throughout this paper, we consider the following Lagrangian of the form:
\begin{equation} \label{Lagrangian}
\mathcal{L}_\phi= K(\phi,X)-\frac{\lambda}{M^3} X\square\phi,~~~~~~~~K(\phi,X)=A(\phi)X+\frac{X^2}{2M^4},
\end{equation} 
where $\lambda >0$ and $M>0$ are parameters and $\square\phi\equiv g^{\mu\nu}\nabla_\mu\nabla_\nu\phi$.
The higher-order kinetic coupling $X\square\phi$ is known as the 
Galileon-type interaction \cite{Nicolis:2008in,Deffayet:2009wt,Deffayet:2009mn}. 
As a working example which allows analytical computations,
we consider $A(\phi)$ given by
\begin{equation} \label{linearized A}
A(\phi)=\left\{
\begin{array}{llll}
-1&\;\;\;\;\;\; & {}&\phi<-\mu,\\
\phi/\mu &\;&{}&-\mu\leq\phi\leq\mu,\\
1&\;&{}&\phi>\mu,
\end{array} 
\right.
\end{equation}
which has shape changes at $\phi=\pm \mu$.
Such an adhoc functional shape of $A(\phi)$ may be justified if $A(\phi)$
is actually determined by another scalar field $\sigma$ as $A(\phi(\sigma))$
and $\sigma$ undergoes a phase transition as $\phi$ evolves to change its amplitude abruptly.
In the same spirit, curvature perturbation in an inflation model with a potential
whose slope changes abruptly has been investigated in \cite{Starobinsky:1992ts}.

For $|\phi| > \mu$, the Lagrangian is invariant under constant shift of $\phi$,
{\it i.e.,} $\phi \to \phi + c$.
Note that the quadratic kinetic interaction $X^2$ can stabilize the theory even 
when $A<0$ \cite{ArkaniHamed:2003uy}. This is similar to the Higgs phenomena where a
quartic self-interaction with positive coupling constant can give
a stable vaccum to a scalar field with negative quadratic mass term.
Here, a phase transition of $\phi$ is led by the change sign of
$A$ from negative to positive. 

The energy-momentum tensor can be divided as $T^\phi_{\mu\nu}=T^K_{\mu\nu}+\frac{\lambda}{M^3} T^G_{\mu\nu}$ with
\begin{eqnarray}
T^K_{\mu\nu}&=&K_X\nabla_\mu\phi\nabla_\nu\phi+g_{\mu\nu}K,\\
T^G_{\mu\nu}&=&g_{\mu\nu}\nabla_\lambda X\nabla^\lambda\phi-2\nabla_{(\mu}X\nabla_{\nu)}\phi
-\square\phi\nabla_\mu\phi\nabla_\nu\phi. 
\end{eqnarray}
 
In terms of the Noether current $J_\mu$, associated with the constant shift transformation $\phi\rightarrow \phi+c$, 
the equation of motion is given by
\begin{eqnarray}\label{EoM}
\nabla_\mu J^\mu &=&K_\phi, \\
J_\mu &=&K_X\nabla_\mu\phi+\frac{\lambda}{M^3}
(\nabla_\mu\nabla_\nu\phi\nabla^\nu\phi-\square\phi\nabla_\mu\phi),
\end{eqnarray}
where $K_\phi\equiv \partial K/\partial\phi$.
On the FLRW background,
we obtain the field energy density and pressure as
\begin{eqnarray}
\label{energy density}
 &&\rho_{\phi}=  2XK_X-K +3\frac{\lambda}{M^3} H\dot{\phi}^3,\\
 &&p_{\phi}=  K-2\frac{\lambda}{M^3} X\ddot{\phi}.\ \ \ \ \
 \label{pressure}
\end{eqnarray}
The equation of motion (\ref{EoM}) of the homogeneous scalar field reads
\begin{equation}\label{eq:eom}
\dot{J}_0+3HJ_0=K_\phi,
\end{equation}
where $J_0=\dot{\phi}(K_X+3\lambda H\dot{\phi}/M^3)$ is the charge density.

\subsection{Dynamics of the spectator field}
We assume that $\phi$ is moving in the shift symmetric phase, {\it i.e.},
$\phi \ll -\mu$ in the inflationary period and reaches transition region $|\phi| < \mu$
in the post-inflationary epoch.
The dynamics in the shift symmetric phase is relatively simple due to vanishing of 
the r.~h.~s. of Eq.~(\ref{eq:eom}).
Since the solution in terms of $J_0$ is given by $J_0 =C/a^3$ ($C:$ integration constant),
$J_0$ quickly approaches zero with the inflationary expansion of the universe.
For simplicity, we assume $J_0=0$ is already achieved to a good approximation
in the inflationary epoch relevant to our scenario.
The equation $J_0=0$ has a trivial solution $\dot{\phi}=0$ as well as a nontrivial solution given by
\begin{equation} \label{eq:shift-symmtry condition}
K_X+3\lambda H\dot{\phi}/M^3=0,
\end{equation} 
for which the energy density \eqref{energy density} reduces to $\rho_\phi=-K$.
Here, we are concerned about the nontrivial solution only, 
and thus the canonical case ($K=X$ and $\lambda=0$) is out of our interest.
Solving Eq.~(\ref{eq:shift-symmtry condition}) with respect to ${\dot \phi}$ yields
\begin{equation}\label{condition of the model}
\dot{\phi}=M\left( -3\lambda H +\sqrt{9\lambda^2 H^2+2M^2}\right),
\end{equation}
where we have picked up only a solution satisfying ${\dot \phi} >0$.

If the Galileon term is subdominant, {\it i.e.},
$\lambda H \ll M$, we have ${\dot \phi} \approx \sqrt{2}M^2$
and $\rho_\phi \approx M^4/2$.
Thus, the spectator field $\phi$ approximately behaves as a cosmological constant.
If, on the other hand, the Galileon term is dominant, {\it i.e.},
$\lambda H \gg M$, we have ${\dot \phi} \approx M^3/(3\lambda H)$
and $\rho_\phi \approx M^6/(18\lambda^2 H^2)$.
Given that $H$ is a decreasing function of time,
$\rho_\phi$ increases with time.
This analysis implies that, 
even if the energy density $\rho_\phi$ is negligible during inflation, 
it finally dominates the Universe if the period of shift symmetric regime
lasts for sufficiently long time.
The dominance of $\phi$ at late time does not happen if $\phi$ 
passes through the transition regime and moves to another shift symmetric
regime $\phi >\mu$ before $\phi$ dominates the Universe.
In order to see this, let us assume $\phi >\mu$.
We will see later that $\phi$ actually becomes larger than $\mu$ in the case
we are interested in.
In this case, solution of Eq.~(\ref{eq:shift-symmtry condition}) becomes
\begin{equation}
\dot{\phi}=M\left( -3\lambda  H \pm \sqrt{9\lambda^2 H^2-2M^2}\right), \label{dot-phi-second-shift}
\end{equation}
which is always negative. 
This means that another solution ${\dot \phi}=0$ of $J_0=0$ is an attractor 
in the current regime.
Given that $J_0 \propto {\dot \phi}$ holds when ${\dot \phi}$ is sufficiently small, 
$J_0 \propto a^{-3}$ means ${\dot \phi}$ approaches zero like ${\dot \phi} \propto a^{-3}$.
In terms of $\rho_\phi$, it scales as $\rho_\phi \propto a^{-6}$.
Thus, once $\phi$ enters the second shift symmetric regime, 
$\rho_\phi$ starts to decay quickly.
Thus, fraction of $\rho_\phi$ to the total energy density takes 
maximum during $\phi$ is in the transition regime.
As we will show, contribution of $\phi$ field perturbation to the curvature perturbation
also takes maximum during the transition regime.
As a result, the curvature perturbation is temporally enhanced,
which may lead to efficient production of primordial black holes.

For the purpose of evaluating how much and how long the curvature perturbation is
temporally enhanced,
we need to analyze the motion of $\phi$ during and after the transition regime.
Given that the Galileon term becomes less important compared to $X^2$ term in $K$
as $H$ decreases (see Eq.~(\ref{condition of the model})),
for simplicity, we consider a case in which the Galileon term has already become
subdominant when $\phi$ reaches $-\mu$.
We do not make particular assumption about the magnitude relation between the Galileon term
and the $X^2$ term during inflation.
Under this setting, we can impose initial conditions such that $\phi=-\mu,~{\dot \phi}=\sqrt{2}M^2$
and $a=1$.
The equation of motion for $\phi$ in our present case ($\phi \ge -\mu$) becomes
\begin{equation}\label{simplified eom}
\frac{d}{dt}
\left[\left(A(\phi)+\frac{\dot{\phi}^2}{2M^4}\right)\dot{\phi}\right]+
3H\left(A(\phi)+\frac{\dot{\phi}^2}{2M^4}\right)\dot{\phi}=
\frac{\dot{\phi}^2}{2}A_\phi(\phi).
\end{equation} 
Meanwhile, the energy density \eqref{energy density} and pressure \eqref{pressure} become
\begin{equation}
\rho_\phi=\frac{\dot{\phi}^2}{2}
\left(A(\phi)+\frac{3\dot{\phi}^2}{4M^4}\right),\;\;\;\;\;\;
p_\phi=\frac{\dot{\phi}^2}{2}
\left(A(\phi)+\frac{\dot{\phi}^2}{4M^4}\right),
\end{equation}
and the Friedmann equation is
\begin{equation}\label{Friedmann eq}
H^2=\frac{1}{3M_P^2}(\rho+\rho_\phi),
\end{equation}
where $\rho$ is the dominant component (descendant of inflation energy density).
If the inflaton has already decayed into radiation, $\rho$ is the radiation energy density.
If it is still in the period of inflaton oscillations, equation
of state of $\rho$ depends on the potential shape.
In our analysis, we assume $\rho$ is in the form of radiation.
Other cases with different values of $w$ can be analyzed similarly.
Note that at the critical time $t=t_c$, where $A(\phi)=\phi=0$,
one finds that $w_\phi=p_\phi/\rho_\phi=1/3$.

To solve the above equation of motion, 
it is convenient to use the dimensionless parameters defined by
\begin{equation}\label{dimensionless parameter}
x\equiv\frac{\phi}{\mu},\;\;\;\;\;
\tau\equiv\sqrt{2}\frac{M^2}{\mu}t,\;\;\;\;\;
h\equiv\frac{1}{a}\frac{da}{d\tau},
\end{equation}
and we denote $x^\prime\equiv dx/d\tau$. 
We choose the origin of time $\tau=0$ when $x =-1$.
We also define the useful dimensionless parameters
$y\equiv \mu/M_{\text{p}}$ and $\epsilon\equiv M^4/\rho_{r,{\rm tra}}$
where $\rho_{r,{\rm tra}}$ is the energy density of radiation 
evaluated at $\tau=0$.

For $-1\leq x\leq 1$, we may convert the
equation of motion \eqref{simplified eom} and
the Friedmann equation \eqref{Friedmann eq} 
into dimensionless forms as
\begin{equation}\label{dimensionless eom}
\frac{d}{d\tau}
\left[\left(x+x^{\prime\,2}\right)x^\prime\right]+
3h\left(x+x^{\prime\,2}\right)x^\prime=
\frac{1}{2}x^{\prime\,2},
\end{equation}
\begin{equation}\label{dimensionless Friedmann eq}
h^2=\frac{y^2}{6}
\left[\frac{1}{\epsilon}\frac{1}{a^4}+\left(x+\frac{3}{2}x^{\prime\,2}\right)x^{\prime\,2}\right].
\end{equation}
Similarly, for $x>1$, Eqs.~\eqref{simplified eom} and
\eqref{Friedmann eq} are given by
\begin{equation}\label{dimensionless eom final}
\frac{d}{d\tau}
\left[\left(1+x^{\prime\,2}\right)x^\prime\right]+
3h\left(1+x^{\prime\,2}\right)x^\prime=0,
\end{equation}
\begin{equation}\label{dimensionless Friedmann eq final}
h^2=\frac{y^2}{6}
\left[\frac{1}{\epsilon}\frac{1}{a^4}+\left(1+\frac{3}{2}x^{\prime\,2}\right)x^{\prime\,2}\right].
\end{equation}
Taking the initial conditions $a=1$, $x=-1$ and $x^\prime=1$ at $\tau=0$,
the equation of state $w_\phi$ for $-1\leq x\leq 1$ is 
\begin{equation}
w_\phi=\frac{2x+x^{\prime\,2}}{2x+3x^{\prime\,2}}.
\end{equation}
This shows $w_\phi=-1$ at $\tau=0$ and $w_\phi=1/3$ when $x=0$. 
Thus, fraction of $\rho_\phi$ becomes maximal exactly when $x=0$.
We can introduce another useful parameter $\eta$ by
\begin{equation}
\eta\equiv \left. \frac{\mu H}{M^2}\right|_{\phi=-\mu}
=\sqrt{\frac{y^2}{3}\left(\frac{1}{\epsilon}+\frac{1}{2}\right)},
\end{equation}
where the second relation is obtained from Eq.~\eqref{dimensionless Friedmann eq} at $\tau=0$.
This quantity measures if $\phi$ passes through the transition regime within Hubble time ($\eta \ll 1$)
or not ($\eta \gg 1$).
When the universe is dominated by radiation, that is $\epsilon a^4\ll 1$,
both Eqs.~\eqref{dimensionless Friedmann eq} and
\eqref{dimensionless Friedmann eq final} become
\begin{equation}
h^2\simeq \frac{y^2}{6\epsilon a^4}\simeq\frac{\eta^2}{2a^4},
\end{equation}
whose solution can be immediately obtained as
\begin{equation}\label{scale factor}
a=(\sqrt{2}\eta\tau+1)^{1/2}.
\end{equation}
By numerically solving the equations of motion above with initial conditions such that
$x(0)=-1,~x'(0)=1,~a(0)=1$, we can determine the time evolution of $\phi$ and $a$.
In the two limiting cases ($\eta \ll 1$ and $\eta \gg 1$), 
analytic computations are possible, which helps us to understand the dynamics qualitatively.
In what follows, we consider the two cases one by one.

\subsubsection{Case I: $\eta\gg 1$}
Let us first consider the case in which $\eta\gg 1$.
In this case, expansion of the Universe cannot be neglected during $\phi$ is in the
transition regime. 
When $0<\tau<1/\eta\ll 1$, we have $h\simeq \eta/\sqrt{2}$,
and the equation of motion \eqref{dimensionless eom} leads to
\begin{equation}
\frac{d}{d\tau}
\left[\left(x+x^{\prime\,2}\right)x^\prime\right]+
\frac{3}{\sqrt{2}}\eta\left(x+x^{\prime\,2}\right)x^\prime=
\frac{1}{2}x^{\prime\,2}.
\end{equation}
During this period, the solution near initial time can be written as
\begin{eqnarray}
x(\tau)&=&-1+\tau-\frac{1}{8}\tau^2+O(\tau^3),\\
x^\prime(\tau)&=&1-\frac{1}{4}\tau+O(\tau^2).
\end{eqnarray}
This implies that we still have $x\simeq-1$ and $x^\prime\simeq 1$
at $\tau=1/\eta$.

\begin{figure}[t]
\begin{center}
\includegraphics[width=7 cm]{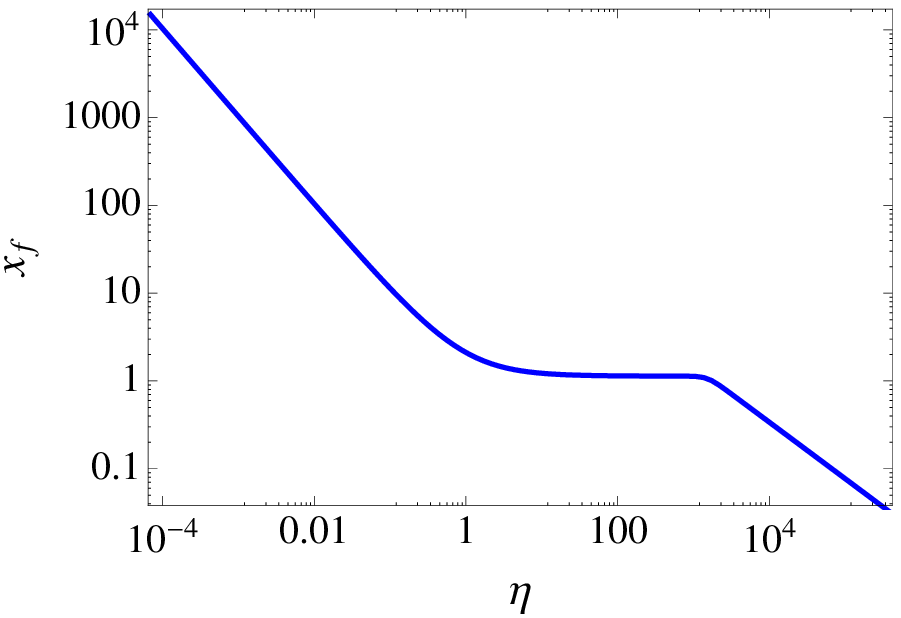}
\includegraphics[width=7 cm]{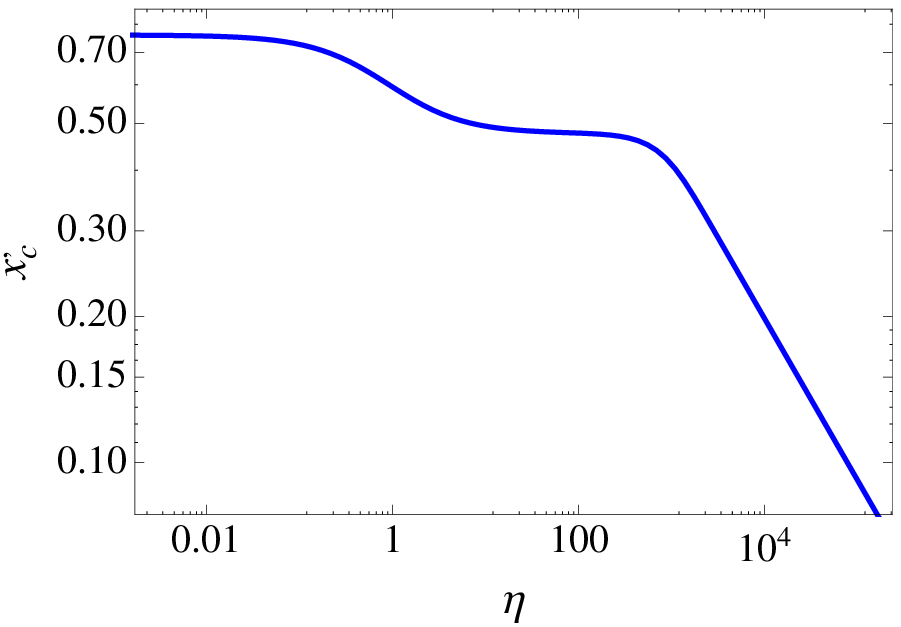}
\end{center}
\caption{The left and right panel shows the value of 
$x_f\equiv x\,(\tau\rightarrow\infty)$ and
$x^\prime_c\equiv x^\prime\,(\tau=\tau_c)$ as a function of $\eta$,
respectively. The initial conditions at $\tau=0$ are $x(0)=-1$,
$x^\prime(0)=1$ and $a(0)=1$, where we use $\epsilon=1\times 10^{-5}$.
In the left panel, $x_f$ shows a plateau in the region
of $1\ll\eta<1/\sqrt{\epsilon}$ and it evolves proportional to $1/\eta$
when $\eta\ll 1$. In the right panel, $x^\prime_c$ shows a plateau in the region of $1\ll\eta<1/\sqrt{\epsilon}$ and $x^\prime_c=3^{-1/4}$
when $\eta\ll 1$. In both figures, the plateau of the value breaks down when
$\eta>1/\sqrt{\epsilon}$.}
\label{Fig 1}
\end{figure}

After $\tau=1/\eta$, we have $h\simeq 1/2\tau$ and Eq.~\eqref{dimensionless eom} becomes
\begin{equation}\label{dimensionless eom large eta}
\frac{d}{d\tau}
\left[\left(x+x^{\prime\,2}\right)x^\prime\right]+
\frac{3}{2\tau}\left(x+x^{\prime\,2}\right)x^\prime=
\frac{1}{2}x^{\prime\,2},
\end{equation}
which shows no explicit dependence on $\eta$. Thus the final value
of $x$ at $\tau\rightarrow \infty$ as well as $x^\prime$ at
$x=0$ are independent of $\eta$.
This is indeed verified by numerical integration of Eq.~\eqref{dimensionless eom}, 
as seen by Fig.~\ref{Fig 1}.
This conclusion is valid as long as the universe remains
dominated by radiation. 
If $\eta$ is too large such that $\epsilon\eta^2\gg 1$, 
one finds $\epsilon a^4\gg 1$ before the critical time $\tau_c$ where $x(\tau_c)=0$. 
In this case the universe is dominated by $\rho_\phi$ before $x=0$ and 
a secondary inflation is induced by the spectator field.

In the case of $\eta<1/\sqrt{\epsilon}$, 
numerical calculation shows that final value of $\phi$ becomes slightly larger than $\mu$
and $\phi$ settles down in the shift symmetric phase. 
Thus, the argument in the paragraph below Eq.~(\ref{dot-phi-second-shift}) can be applied 
and $\phi$ field eventually evolves as a free scalar with $w_\phi=1$.
Keeping in mind that $w_\phi(0)=-1$ and $w_\phi(\tau_c)=1/3$,
the energy density $\rho_\phi$ is decaying slower (faster) than 
$\rho_r$ before (after) the critical time and $r_\phi$ reaches
a maximum value $r_\phi^{\text{max}}\simeq \epsilon\eta^2$ around $\tau_c$.
 
\subsubsection{Case II: $\eta\ll 1$}
In the opposite case where $\eta\ll 1$, the effect of the
cosmic expansion is negligible during $\phi$ moves from $-\mu$ to $\mu$. 
Thus we can drop the Hubble friction terms in
Eq.~\eqref{dimensionless eom}, which yields
\begin{equation}
\frac{d}{d\tau}\left[\left(x+x^{\prime\,2}\right)x^\prime\right]
=\frac{1}{2}x^{\prime\,2}.
\end{equation}
This leads to the conservation of energy density as
\begin{equation}\label{density conservation}
x^{\prime\,2}\left(x+\frac{3}{2}x^{\prime\,2}\right)=\frac{1}{2},
\end{equation}
where we have used the initial conditions $x=-1$ and $x^\prime=1$
at $\tau=0$. It is easy to obtain from Eq.~\eqref{density conservation}
that $x^\prime=3^{-1/4}$ at $x=0$ and that $x^\prime=1/\sqrt{3}$ at $x=1$.

We may now solve the dynamics for $x>1$. 
Taking $\tau_1$ as the time when $x(\tau_1)=1$, we have the initial value 
$x^\prime(\tau_1)=1/\sqrt{3}$. Since $\epsilon\eta^2\ll 1$ always holds
when $\eta\ll 1$, the universe keeps radiation-dominated and the
scale factor follows Eq.~\eqref{scale factor}. 
Meanwhile, by integrating once Eq.~\eqref{dimensionless eom final}, we have
\begin{equation}\label{solution for x>1}
a^3\left(1+x^{\prime\,2}\right)x^\prime=\frac{4}{3\sqrt{3}},
\end{equation}
where the initial conditions $x^\prime(\tau_1)=1/\sqrt{3}$ and
$a(\tau_1)=1$ are used. Here, it suffices to approximate
$1+x^{\prime\,2}\rightarrow 1$ since $x^\prime\rightarrow 0$ is the
attractor of this phase, and thus $a^3x^\prime\simeq 4/3\sqrt{3}$
after some time.
With this approximation, we can evaluate the final value of $x$ as
\begin{equation}\label{xf}
x_f=1+\int^\infty_{\tau_1}\,d\tau\, \frac{4}{3\sqrt{3}\, a^3(\tau)} \approx \frac{4\sqrt{2}}{3\sqrt{3} \eta}.
\end{equation}  
Using Eq.~\eqref{scale factor}, 
we find that $x_f\propto 1/\eta$ in the present case, which is confirmed 
by the numerical integration of the equations of motion (see the left panel of Fig.~\ref{Fig 1}).

\section{Generation of the Curvature perturbation}
\label{cur}
In this section, we will show that the curvature perturbation is temporally enhanced
at around a critial time when $\phi=0$.
Since such curvature perturbation is sourced by the perturbation of $\phi$,
we will first compute the $\phi$ field perturbation and then proceed to 
evaluate the curvature perturbation by means of the $\delta N$ formalism.

\subsection{Spectator field perturbation}
To study the perturbation of the spectator field, 
we first decompose the scalar field into a homogeneous and an inhomogeneuos parts:  
\begin{equation}\label{field decomposition}
\phi(t,\textbf{x})\rightarrow\phi(t)+\delta\phi(t,\textbf{x}).
\end{equation}
Given the subdominance of $\phi$, we treat $\phi$ as a test field on a fixed background
spacetime (de Sitter space).
 
Substituting the decomposition Eq.~\eqref{field decomposition} into the 
original Lagrangian \eqref{Lagrangian} and expanding it up to second order in perturbation,
in the shift symmetric phase,
we have
\begin{equation}
{\cal L}_\phi^{(2)}= \int dt d^3x~\bigg[ \frac{a^3}{2} {\cal D} {\dot {\delta \phi}}^2-\frac{a}{2}{\cal C} 
{\left( {\vec \nabla} \delta \phi \right)}^2 \bigg], \label{2nd-Lag}
\end{equation}
where ${\cal C}$ and ${\cal D}$ are given by \cite{Wang:2011dt},
\begin{equation}
{\cal C}=K_X+2\frac{\lambda}{M^3} ( {\ddot \phi}+2H{\dot \phi}),
~~~~~{\cal D}=K_X+2XK_{XX}+6\frac{\lambda}{M^3} H{\dot \phi}.
\end{equation}
Variation of Eq.~(\ref{2nd-Lag}) yields
\begin{equation}\label{eom of perturbation}
\delta\ddot{\phi}+\left( 3+\frac{\dot{\cal D}}{H{\cal D}}\right)H\delta\dot{\phi}-\frac{c_s^2}{a^2}\nabla^2\delta\phi=0,
\end{equation}
where $c_s^2={\cal C}/{\cal D}$ is the sound speed squared.
To avoid instablities, one requires $c^2_s\geq 0$.
As we will see later, this condition is satisfied in the situation we consider.
On the nearly de Sitter background, we have $|{\dot {\cal D}}/(H{\cal D})| \ll 1$ and
the dominant friction term takes the standard form.
Thus, for any mode outside the sound horizon ($c_s k<aH$), two basic solutions of Eq.~(\ref{eom of perturbation})
are constant mode and the decaying mode.
The former determines the amplitude of $\delta \phi$ at the time when $\phi$ reaches $-\mu$. 
Quantization of the second order Lagrangian on the nearly de Sitter background gives power spectrum of $\delta \phi$ 
on large scales $(c_s k \ll aH)$ as
\begin{equation}
{\cal P}_\phi = \frac{H_*^2}{4\pi^2 c_s^3 {\cal D}},
\end{equation}
where the subscript ${}_*$ means that corresponding quantity is evaluated at a time when the relevant mode
crosses the sound horizon $c_s k=a_* H_*$.

If the Galileon term is negligible at the time of sound horizon crossing, {\it i.e.},
$M \gg \lambda H_*$, then we have
\begin{equation}
{\cal C} \approx 4\sqrt{2} \frac{\lambda H_*}{M},~~~~~{\cal D} \approx 2,~~~~~c_s^2 \approx 4\sqrt{2} \frac{\lambda H_*}{M},
~~~~~{\cal P}_\phi \approx \frac{1}{32\sqrt{2}} {\left( \frac{H_*}{2\pi} \right)}^2 {\left( \frac{\lambda H_*}{M} \right)}^{-3/2}. \label{P-sub-gali}
\end{equation}
It is evident that one can obtain a large spectrum from a very small 
sound speed by taking $\lambda\rightarrow 0$.
If, on the other hand, the Galileon term is dominant at the time of sound horizon crossing, {\it i.e.},
$M \ll \lambda H_*$, then we have
\begin{equation}
{\cal C} \approx \frac{1}{3},~~~~~{\cal D} \approx 1,~~~~~c_s^2 \approx \frac{1}{3},
~~~~~{\cal P}_\phi \approx 3\sqrt{3}{\left( \frac{H_*}{2\pi} \right)}^2.
\end{equation}

One should also consider the stochastic effect of the quantum fluctuations of the spectator field, 
$\Delta\phi_{\text{quant}}\sim H_\ast/(2\pi\sqrt{D}c_s^{3/2})$, during inflation. 
Since each quantum $k$ mode of ${\dot \phi}$ contributes to the background ${\dot \phi}$ at 
each sound horizon crossing time, 
background ${\dot \phi}$ changes on average by the amount of 
$\Delta {\dot \phi}_{\text{quant}}=H \Delta \phi_{\text{quant}}$ at each Hubble time.
If this variation of ${\dot \phi}$ exceeds a certain threshold which we denote by $\Delta {\dot \phi}_{\rm cl}$,
${\dot \phi}$ of the attractor solution Eq.~(\ref{condition of the model}) can jump
into other attractor solutions. 

In the case the Galileon is subdominant ($M \gg \lambda H_*$),
the positive solution of Eq.~(\ref{condition of the model}) is approximately equal
to the absolute value of the negative solution.
If transitions to both ${\dot \phi}=0$ and negative ${\dot \phi}$ could happen in the Hubble time,
the observable Universe when inflation ends would consist of three kinds of patches corresponding
to three different attractor solutions.
Recalling that the negative ${\dot \phi}$ behaves as a cosmological constant, 
the negative ${\dot \phi}$ patch is eventually dominated by $\phi$ 
and the second (eternal) inflation by $\phi$ commences.
In addition to this, the background solution ${\dot \phi}=0$ is unstable
since the fluctuation around this solution becomes ghost. 
We require that such transitions to other attractor solutions do not happen in all of the sound horizon patches
($\sim {(c_s^{-1} e^{N_{\rm inf}})}^3=e^{-3 \ln c_s +3N_{\rm inf}}$) constituting the observable Universe,
where $N_{\rm inf}$ is the number of $e$-folds of inflation covering the observable Universe.
In other words, probability of jumping of ${\dot \phi}$ to ${\dot \phi}=0$ must be 
suppressed by more than $\sqrt{6(N_{\rm inf}- \ln c_s)}~ \sigma$ Gaussian probability;
\begin{equation}
H \Delta \phi_{\text{quant}} < \frac{1}{\sqrt{6(N_{\rm inf}- \ln c_s)}} \Delta {\dot \phi}_{\rm cl}. \label{con0}
\end{equation}
Given our ignorance of the precise value of $N_{\rm inf}$,
we assume $N_{\rm in}-\ln c_s =60$, for definiteness.
Our condition (see Eq.~(\ref{con1})) changes only little even if we 
choose another value as long as it is ${\cal O}(50-70)$.
Using Eq.~(\ref{P-sub-gali}) and $\Delta {\dot \phi}_{\rm cl}=\sqrt{2}(1-\frac{1}{\sqrt{3}})M^2$, 
the condition (\ref{con0}) leads to
\begin{equation}
\frac{H_*}{M} < {\left( \frac{4\times2^{3/4}}{3\sqrt{5}} 
\left( 1-\frac{1}{\sqrt{3}}\right) \right)}^{4/5} \lambda^{3/5} \approx 0.5 \lambda^{3/5},~~~~~{\rm for}~~M \gg \lambda H_*. \label{con1}\\
\end{equation}

On the other hand, in the Galileon dominant case ($M \gg \lambda H_*$),
since the positive solution of Eq.~(\ref{condition of the model}) is much smaller than
the absolute value of the negative solution,
transition only to ${\dot \phi}=0$ occurs unless the Hubble parameter is higher
than $\max \{\lambda M, \lambda^{-1}M \}$.
As stated above, the background solution ${\dot \phi}=0$ is unstable quantum mechanically. 
To avoid this situation,
we impose again Eq.~(\ref{con0}), which leads to
\begin{equation}
\frac{H_*}{M} < {\left( \frac{\pi}{54 \times 3^{3/4}} \right)}^{1/3} \lambda^{-1/3}
\approx 0.3 \lambda^{-1/3},~~~~~{\rm for}~~M \ll \lambda H_*.
\label{con2} 
\end{equation}

\begin{figure}[t]
\begin{center}
\includegraphics[width=10cm]{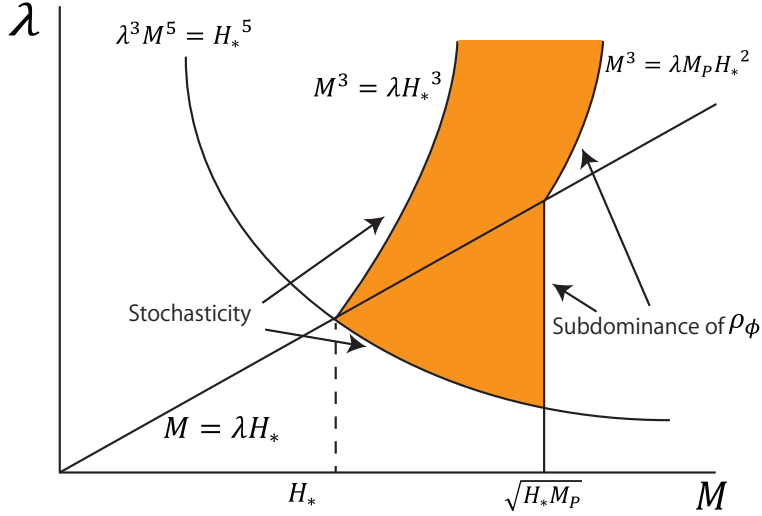}
\end{center}
\caption{The region satisfying conditions (\ref{con1})-(\ref{con4}) is colored orange.}
\label{allowed}
\end{figure}

Final condition imposed for the model parameters is that $\rho_\phi$ should not exceed the
inflaton energy density.
This condition leads to
\begin{eqnarray}
&&M < 6^{1/4} \sqrt{M_P H_*},~~~~~{\rm for}~~M \gg \lambda H_*, \label{con3}\\
&&M < 2^{1/6}3^{1/2} \lambda^{1/3} M_P^{1/3} H_*^{2/3},~~~~~{\rm for}~~M \ll \lambda H_*. \label{con4} 
\end{eqnarray}
The region satisfying these conditions (\ref{con1})-(\ref{con4}) is shown in Fig.~\ref{allowed}.
We find that $M>H_*$ is a necessary condition to satisfy the conditions.

\subsection{Curvature perturbation}
Having understood the amplitude of the spectator field perturbation generated 
during inflation,
we can now study evolution of the curvature perturbation sourced by the $\phi$ field
during transition regime.
To this end, we adopt the $\delta N$ formalism \cite{Starobinsky:1986fxa,Sasaki:1995aw,Nambu:1997wh,Sasaki:1998ug,Lyth:2004gb} to compute the 
curvature perturbation $\zeta$ defined for the total energy density.
On super-Hubble scales, the metric can be written as
\begin{equation}
ds^2 =-\alpha^2 dt^2+a^2(t) e^{2\psi} \delta_{ij}dx^i dx^j.
\end{equation}
In this form, $\psi$ represents the curvature perturbation on the $t={\rm const.}$
hypersurface.
If we choose the $t={\rm const.}$ hypersurface at an initial time (in our present case,
we take it to be the time when $\phi$ reaches $-\mu$.) to be flat slicing
and at later time of interest to be constant energy density slicing,
then, denoting the curvature perturbation on such slicing as $\zeta$, 
$\zeta$ at any point is equal to the perturbation of number of $e$-folds computed
as if the same point were evolving like the unperturbed FLRW universe with initial
conditions specified by the sourcing field perturbations evaluated on the initial 
flat hypersurface.

Having briefly reviewed the basics of the $\delta N$ formalism, 
let us now formulate how to compute $\zeta$ in our current situation.
Due to the fronzen-in of $\phi$ field perturbation on super-Hubble scales
in the shift symmetric regime,
perturbation of $\phi$ on the initial flat hypersurface(time when $\phi=-\mu$) 
is nothing but $\delta\phi_\ast$ given in the last subsection.
If there is perturbation $\delta\phi_\ast$ at a given position, then
the time $\phi$ arrives at $-\mu$ deviates from the one in the unperturbed
universe by $\delta t_{\rm tra}=\delta\phi_\ast/\dot{\phi}$. In terms of the
\textit{e}-fold number, the corresponding deviation is
\begin{equation}
\delta N_{\rm tra}=H_{\rm tra}\delta t_{\rm tra}=\frac{H_{\rm tra} \delta\phi_\ast}{\dot{\phi}}.
\end{equation}
This results in some difference of the initial radiation energy density
evaluated at $\phi=-\mu$ by an amount
\begin{equation}
\delta\rho_{r,{\rm tra}}=\rho_{r,{\rm tra}}e^{4\delta N_{\rm tra}}-\rho_{r,{\rm tra}}
\approx 4\rho_{r,{\rm tra}}\delta N_{\rm tra}.
\end{equation}

At each position, the perturbation $\delta\rho_{r\ast}$ changes the
parameter $\epsilon$ by
\begin{equation}
\delta\epsilon=\frac{M^4}{\rho_{r,{\rm tra}}+\delta\rho_{r,{\rm tra}}}
-\frac{M^4}{\rho_{r,{\rm tra}}}\approx -4\epsilon\delta N_{\rm tra}
=-4\epsilon\,\frac{\eta\,\delta x_\ast}{\sqrt{2}},
\end{equation}
where $\delta x_\ast\equiv\delta\phi_\ast/\mu$.
Let us denote by $N_f(\epsilon,y)$ the number of \textit{e}-folds measured
from the initial time when $x=-1$, $x^\prime=1$ to the time when
$h=h_f$ for given parameters $(\epsilon,y)$. 
Then, at each perturbed position,
the total number of \textit{e}-folds measured form the initial flat
hypersurface to the time when the Hubble parameter decreases to a fixed
value $H_f$ is a sum of $-\delta N_{\rm tra}$ and 
$N_f(\epsilon+\delta\epsilon,y)$. 
Identifying this amount of $\delta N$ as $\zeta$,
we conclude that
\begin{equation}\label{delta N formalism}
\zeta=-\delta N_{\rm tra}+
\left.\frac{\partial N_f}{\partial\epsilon}\right|_y\delta\epsilon
=-\eta
\left(1+4\epsilon\left.\frac{\partial N_f}{\partial\epsilon}\right|_y\right)
\frac{\delta x_\ast}{\sqrt{2}}.
\end{equation}
As a result, we can immediately compute $\zeta$ by using the above formula
once $N_f(\epsilon,y)$ is known. 
In the present analysis, contribution of inflaton perturbation to $\zeta$,
which is assumed to be responsible for the observed CMB anisotropies,
is neglected since it does not play any role in temporally enhancing the curvature
perturbation.
In what follows we analytically evaluate the curvature perturbation in the 
limiting cases of $\eta\gg 1$ and $\eta\ll 1$.

\subsubsection{Case I: $\eta\gg 1$} 
In this case, the scale factor during the radiation domination is given by
Eq.~\eqref{scale factor}, and the Hubble parameter quickly becomes 
$h\approx1/(2\tau)$ after $\tau=1/\eta\ll 1$. When $\tau>1/\eta$, the dynamics
of $x$ follows 
\begin{equation}
\frac{d}{d\tau}
\left[\left(A(x)+x^{\prime\,2}\right)x^\prime\right]+
\frac{3}{2\tau}\left(A(x)+x^{\prime\,2}\right)x^\prime=
\frac{1}{2}x^{\prime\,2}A_x,
\end{equation}
where $A(x)=x$ for $-1\leq x\leq 1$ and $A(x)=1$ for $x>1$. 
Since $x(\tau)$ is independent of $\eta$, we may express the dimensionless
energy density as a function of time $\tau$ as
\begin{equation}
x^{\prime\,2}\left(A(x)+\frac{3}{2}x^{\prime\,2}\right)=f(\tau)=
f\left(\frac{a^2}{\sqrt{2}\,\eta}\right).
\end{equation}
Then Friedmann equation \eqref{dimensionless Friedmann eq} becomes
\begin{equation}
h_f^2=\frac{y^2}{6}
\left[\frac{1}{\epsilon e^{4N_f}}+f\left(\frac{e^{2N_f}}{\sqrt{2}\,\eta}\right)\right],
\end{equation}
from which we obtain
\begin{equation}
4\epsilon\left.\frac{\partial N_f}{\partial\epsilon}\right|_y
=-1+\epsilon\,\eta^2\tau^3f^\prime(\tau).
\end{equation}
By using the formula \eqref{delta N formalism}, 
the curvature perturbation reads
\begin{equation}\label{zeta for large eta}
\zeta=-\epsilon\,\eta^3\tau^3f^\prime(\tau) \frac{\delta x_\ast}{\sqrt{2}}.
\end{equation}
Given that $\tau$ and $f(\tau)$ are $\mathcal{O}(1)$ quantities at around
$x=0$, we find that $\zeta/\delta x_\ast$ takes an amplitude of 
$\mathcal{O}(\epsilon\eta^3)$ for $\tau\simeq\tau_c$.
In the case of $1\ll\eta<1/\sqrt{\epsilon}$, the dynamics eventually enters 
the second shift symmetric phase($x>1$) and $\phi$ starts to behave as a 
free scalar field $x^\prime\sim 1/a^3$ and $f(\tau)\sim e^{-6N_f}$.
In this phase, $\zeta$ decays in proportion to $a^{-2}$.
Thus, $\zeta$ is temporally enhanced in the transition period.
Numerical integration of the background equations of motion shows $\tau_c \approx 1.4$
and $f'(\tau_c) \approx -0.1$.
Thus, we have $\zeta (\tau_c)/\delta x_* \approx 0.2 ~\epsilon \eta^3$,
which basically gives the maximal amplitude of the enhanced curvature perturbation.

\subsubsection{Case II: $\eta\ll 1$}
In this regime, the universe expands little from $x=-1$ to $x=1$.
Therefore we may evaluate $N_f$ only after $x=1$.
Recalling that, for $x>1$, we have obtained a solution from Eq.~\eqref{solution for x>1} as
\begin{equation}
x^\prime\simeq \frac{4}{3\sqrt{3}}\frac{1}{a^3},
\end{equation}
the Friedmann equation \eqref{dimensionless Friedmann eq final} can be
written as
\begin{equation}
h_f^2=\frac{y^2}{6}
\left(\frac{1}{\epsilon e^{4N_f}}+\frac{16}{27}\frac{1}{e^{6N_f}}\right).
\end{equation}
Since $\epsilon\ll 1$, one finds that $N_f(\epsilon,y)$ is given by
\begin{equation}
-4N_f\approx\ln\left(\frac{6h_f^2\epsilon}{y^2}\right)
\left[1-\frac{8}{81}\frac{y^2}{h_f^2}
\left(\frac{6h_f^2\epsilon}{y^2}\right)^{3/2}\right].
\end{equation}
Plugging this result into Eq.~\eqref{delta N formalism}, we finally arrive at
\begin{equation}
\zeta=\frac{16}{27}\frac{\epsilon\eta}{e^{2N_f}}(6N_f-1) \frac{\delta x_\ast}{\sqrt{2}}.
\end{equation}
It is evident that the curvature perturbation $\zeta/\delta x_\ast$ 
has a maximal amplitude of $\frac{16}{9\sqrt{2}} e^{-4/3}\epsilon\eta$ when 
$N_f= 2/3$ and gradually decays in proportion to $1/a^2$ after that.

\section{Primordial black holes and reheating of the Universe}
\label{pbhs}
It has been pointed out that temporal enhancement of the primordial curvature perturbation 
can leave imprints through the black hole formation at the time the enhanced curvature perturbation 
reenters the horizon \cite{Suyama:2011pu}.
The same consequence is expected in our present scenario, which we investigate in this section.

Primordial black hole are formed when curvature perturbation with an amplitude 
close to unity enters the Hubble radius during radiation domination with their
typical mass given by the horizon mass at that time \footnote{PBH mass
can be significantly smaller than the horizon mass if the perturbation amplitude is
very close to the threshold value \cite{Yokoyama:1998xd, Niemeyer:1999ak}. 
We do not consider this effect in this paper.}.
Precise threshold amplitude of $\zeta$ depends on the initial 
perturbation profile \cite{Shibata:1999zs,Nakama:2013ica}. 
In \cite{Nakama:2013ica}, 
it was found that threshold value of the comoving density perturbation smoothed 
over the Hubble horizon at the time of horizon crossing required 
for the PBH formation ranges from $0.45 \sim 0.5$.
In terms of $\zeta$, the corresponding range is $0.59 \sim 0.63$.
Since our purpose is to identify parameter region where PBHs are produced
abundantly, it is sufficient to take a single threshold value, say $\zeta_{\rm th}=0.6$,
to estimate the abundance of PBHs.
Approximating the statistical distribution as Gaussian, the fraction
of the Universe collapsing into PBHs is estimated as
\begin{equation}
\beta = \int_{\zeta_{\rm th}}^\infty~\frac{1}{\sqrt{2\pi} \zeta (\tau_c)}~\exp \left( -\frac{\zeta'^2}{2\zeta^2 (\tau_c)} \right) d\zeta'
\simeq  \frac{\zeta (\tau_c)}{\sqrt{2\pi} \zeta_{\rm th}} 
\exp \left( -\frac{\zeta_{\rm th}^2}{2\zeta^2 (\tau_c)} \right)
\approx 0.7 \zeta(\tau_c) e^{-\frac{0.18}{\zeta^2(\tau_c)}}.
\end{equation}

Since the amplitude of the enhanced $\zeta$ depends on $\epsilon$ (see last section),
let us first evaluate possible range of $\epsilon$ to be consistent with the recent measurement
of the tensor mode by BICEP2 \cite{Ade:2014xna}.
Detection of the tensor mode with the tensor-scalar ratio $r={\cal O}(0.1)$ directly
gives the Hubble parameter $H_*$ during inflation as
\begin{equation}
H_* \simeq 8 \times 10^{13}~{\rm GeV}~{\left( \frac{r}{0.1} \right)}^{1/2}.
\end{equation}
We use $H_* = 8 \times 10^{13}~{\rm GeV}$ as a reference value of $H_*$ hereafter.
Cases in which $H_*$ is smaller than this value can be studied in a similar way.
Since the total energy density during inflation is larger than the one in the post-inflationary era,
we have
\begin{equation}
\epsilon > \frac{M^4}{3M_P^2 H_*^2} > \frac{H_*^2}{3M_P^2}=4\times 10^{-10}, \label{low-e}
\end{equation}
where we have used an inequality $M>H_*$ obtained in the last section to obtain the second inequality.
This is an absolute lower bound on $\epsilon$.
If we consider realistic situation, 
since the energy density when $\phi=-\mu$ is smaller than
that in the inflationary epoch, the actual lower bound on $\epsilon$ becomes larger than 
Eq.~(\ref{low-e}).
To estimate it, let us assume that the Hubble parameter at the end of inflation is given by
$H_{\rm end}=H_*/\sqrt{N}$, where $N$ is the number of $e$-folds during inflation since
the CMB scale left the Hubble horizon.
This relation holds in the chaotic inflation model with quadratic inflaton potential and
also, up to ${\cal O}(1)$ factor difference, in other polynomial potentials.
As the most extreme case, we consider that termination of inflation is immediately
followed by reheating (instant reheating) and $\phi$ comes to $-\mu$ at that time.
In such a case, we have
\begin{eqnarray}
&&\epsilon = \frac{N M^4}{3M_P^2 H_*^2} = 2\times 10^{-4} \left( \frac{N}{50}\right) {\left( \frac{M}{10 H_*} \right)}^4 {\left( \frac{H_*}{8\times 10^{13}~{\rm GeV}}\right)}^2, \\
&&\eta=4~\left( \frac{\mu}{0.1 M_P} \right) {\left( \frac{N}{50} \right)}^{-1/2} {\left( \frac{M}{10H_*} \right)}^{-2}{\left( \frac{H_*}{8\times 10^{13}~{\rm GeV}}\right)}^{-1}. \label{ep-eta}
\end{eqnarray}
The later $\phi$ arrives at $-\mu$, the larger $\epsilon$ becomes.
This estimate shows that there is parameter space in which $\epsilon$ remains less than unity and 
the second inflation by $\phi$ field is circumvented.

With $\epsilon$ and $\eta$ given by Eqs.~(\ref{ep-eta}), 
let us first evaluate $\zeta (\tau_c)$ in the Galileon dominant case ($\lambda H_* \gg M$)
for completeness.
In this case, we find
\begin{equation}
\zeta (\tau_c)=
  \begin{cases}
    3\times 10^{-8} {\left( \frac{N}{50} \right)}^{1/2} {\left( \frac{M}{10 H_*} \right)}^2 
    {\left( \frac{H_*}{8\times 10^{13}~{\rm GeV}}\right)}^2~~~{\rm for}~~ \eta \ll 1, \\
    3 \times 10^{-7} {\left( \frac{\mu}{0.1 M_P} \right)}^2 {\left( \frac{N}{50} \right)}^{-1/2}
    {\left( \frac{M}{10 H_*} \right)}^{-2}~~~{\rm for}~~ \eta \gg 1.
  \end{cases}
\end{equation}
Given that $M \lesssim 100~H_*$ and $\mu \lesssim M_P$ to avoid the second inflation by $\phi$
as well as $M>H_*$ obtained in the last section, we find that
$\zeta (\tau_c)$ is significantly lower than unity in both cases.
Therefore, there is no observational trace if the Galileon term dominates during inflation.

Hereafter, we consider the opposite case ($M \gg \lambda H_*$)
for which we described background evolution in some detail in Sec.~\ref{model}.
In this case, we have
\begin{equation}
\zeta (\tau_c)=
  \begin{cases}
    10^{-8} {\left( \frac{N}{50} \right)}^{1/2} {\left( \frac{M}{10 H_*} \right)}^{11/4} 
    {\left( \frac{H_*}{8\times 10^{13}~{\rm GeV}} \right)}^2\lambda^{-3/4}~~~{\rm for}~~ \eta \ll 1, \\
    10^{-7} {\left( \frac{\mu}{0.1 M_P} \right)}^2 {\left( \frac{N}{50} \right)}^{-1/2}
    {\left( \frac{M}{10 H_*} \right)}^{-5/4} \lambda^{-3/4}~~~{\rm for}~~ \eta \gg 1.
  \end{cases}
\end{equation}
Due to negative power dependence on $\lambda$, amplitude of the curvature perturbation becomes
as large as unity for $\lambda \ll 1$.
Notice that $\lambda < 1$ automatically satisfies $M \gg \lambda H_*$ 
since $M > H_*$ is always imposed.
For instance, taking $\mu=M_P$ in the latter case, 
which is the possible maximal value of $\mu$ to avoid the second inflation by $\phi$, 
$\zeta (\tau_c)$ exceeds unity for $\lambda \lesssim 2\times 10^{-7}$.
For such a small value of $\lambda$, PBHs would be overproduced,
and $\lambda$ should take a larger value.
Even if $\lambda$ is larger than this value, PBHs are still produced (but not as efficiently
as the case of smaller $\lambda$) since $\zeta >1$ is realized at the tail of the probability 
distribution of $\zeta$.
As mentioned before, typical mass of PBHs is given by the horizon mass at the time of PBH formation as
\begin{equation}
M_{\rm PBH} =\frac{1}{2G H_c} \approx 100~{\rm g}~{\left( \frac{M}{10~H_*} \right)}^{-2}
\left( \frac{\mu}{0.1~M_P} \right) 
{\left( \frac{H_*}{8\times 10^{13}~{\rm GeV}}\right)}^{-2}. \label{pbhmass}
\end{equation}
These BHs evaporate by the Hawking radiation later.
The evaporation time $t_{\rm ev}$ is given by
\begin{equation}
t_{\rm ev}=4\times 10^{-22}~{\rm s} ~{\left( \frac{g_{\rm rel}}{200} \right)}^{-1}
{\left( \frac{M_{\rm PBH}}{100~{\rm g}} \right)}^3,
\end{equation}
where $g_{\rm rel}$ is the degrees of freedom of relativistic particles that constitute radiation.
The corresponding Hubble parameter and the radiation temperature are given by
\begin{equation}
H_{\rm ev} \approx 8\times 10^{-4}~{\rm GeV}~\left( \frac{g_{\rm rel}}{200} \right)
{\left( \frac{M_{\rm PBH}}{100~{\rm g}} \right)}^{-3},
~~~~T_{\rm ev} \approx 2 \times 10^7~{\rm GeV}
{\left( \frac{g_{\rm rel}}{200} \right)}^{1/4} {\left( \frac{M_{\rm PBH}}{100~{\rm g}} \right)}^{-3/2}.
\end{equation}
The value of $\beta$ for which the PBHs just start being the dominant component at the
time of their evaporation is determined as
\begin{equation}
\beta = \sqrt{\frac{H_{\rm ev}}{H_c}} \approx 2\times 10^{-8}~
{\left( \frac{g_{\rm rel}}{200} \right)}^{1/2}{\left( \frac{M_{\rm PBH}}{100~{\rm g}} \right)}^{-1}.
\end{equation}
The corresponding $\zeta (\tau_c)$ for $\beta \simeq 2\times 10^{-8}$ is given by $\zeta (\tau_c) \approx 0.11$.
To conclude, for $\zeta (\tau_c) \approx 0.11$, PBHs would dominate the Universe before they evaporate
and the Universe is again reheated by the evaporation of PBHs.

\begin{figure}[t]
\begin{center}
\includegraphics[width=10 cm]{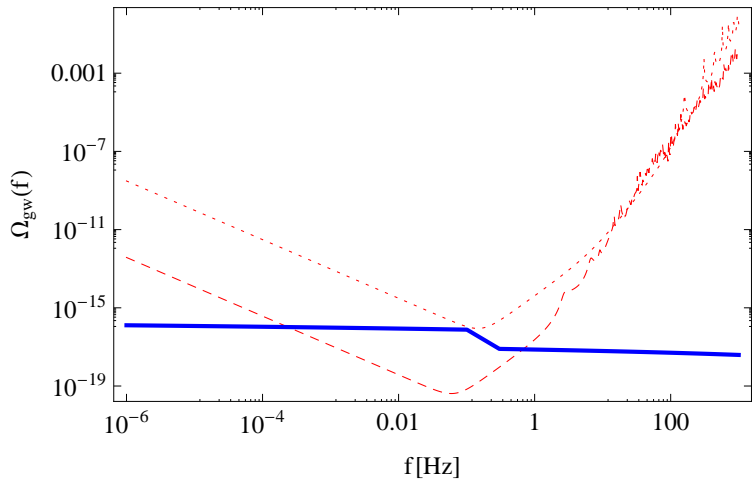}
\end{center}
\caption{Present primordial tensor mode spectrum for $f_{\rm low}=0.1~{\rm Hz}$
and $f_{\rm high}=0.3~{\rm Hz}$ (thick blue curve).
To draw the curve, we have assumed the chaotic inflation by quadratic inflaton potential
for which $r\simeq 0.15$ at the CMB scale.
Dotted(dashed) red curve is the sensitivity curve of the correlated analysis of DECIGO
(ultimate-DECIGO), assuming 10 year observation.
}
\label{gws}
\end{figure}

The scenario of reheating by PBH evaporation in our model can be tested 
by the future space interferometer DECIGO \cite{Seto:2001qf}.
After the PBHs dominate the Universe, the Universe undergoes the matter dominance expansion $a \sim t^{2/3}$
until the PBHs evaporate.
During the PBH dominance, the tensor modes of inflationary origin are diluted like $\propto a^{-1}$
compared to the PBH energy density.
This dilution stops after the PBH evaporation and the Universe undergoes the radiation dominance expansion.
As a result, the spectrum of tensor modes today ($\Omega_{\rm gw}$) exhibits red-tilt in a frequency range corresponding
to the period of PBH dominance like $\Omega_{\rm gw} \propto f^{-2}$, 
while $\Omega_{\rm gw} \propto f^0$ outside that range \cite{Seto:2003kc}.
In terms of the today's frequency, two frequencies at the edges of this range are given by
\begin{equation}
f_{\rm low} \approx 0.8~{\rm Hz} ~{\left( \frac{g_{\rm rel}}{200} \right)}^{1/6} \left( \frac{T_{\rm ev}}{2\times 10^7~{\rm GeV}}\right),
~~~~~f_{\rm high} \approx {\left( \frac{\beta}{2\times 10^{-8}} \right)}^{2/3} f_{\rm low}.
\end{equation}
Amplitude of $\Omega_{\rm gw}$ for $f>f_{\rm high}$ is suppressed by a factor ${(f_{\rm low}/f_{\rm high})}^2$
compared with that for $f<f_{\rm low}$.
In Fig.~\ref{gws}, we show one example of the resultant tensor mode spectrum with sensitivity curves of DECIGO
and ultimate-DECIGO \cite{Kudoh:2005as, sachikuro}.
For the computation of the present tensor mode spectrum over the DECIGO band,
see \cite{Nakayama:2008ip,Nakayama:2008wy}.
We find that ultimate-DECIGO can measure the tensor spectrum in a range 
covering both $f_{\rm low}$ and $f_{\rm high}$ if both quantities take suitable values.

So far, we have fixed the Hubble parameter during inflation as the one suggested by the BICEP2 measurement.
If the Hubble parameter is smaller than this value, 
temporal enhancement happens at lower energy scales and the mass of the PBHs becomes larger. 
For instance, if we take $H_* \approx 10^{-4}~{\rm GeV}$,
Eq.~(\ref{pbhmass}) gives $M_{\rm PBH} \approx 10^4~M_\odot$.
Evaporation time of such PBHs is much longer than the age of the Universe,
and reheating by BH evaporation does not work. 
Yet, those PBHs are interesting in the context of supermassive BHs residing in present galaxies
as PBHs in the mass range $10^3 M_\odot \sim 10^5M_\odot$ can be seeds for the 
present supermassive BHs \cite{Bean:2002kx}.
 
\section{conclusion}
\label{conc}
In this paper, we have studied a model in which the inflaton-generated curvature perturbation 
explains the observed CMB temperature anisotropies but primordial curvature perturbation is 
temporarily dominated in the primordial era by another 
perturbation converted from an evolving spectator field whose energy density is negligible during inflation.

As a working model, we have considered the spectator field whose Lagrangian has, in addtion to the canonical kinetic
term, a $X^2$ term ($X=-\frac{1}{2}{(\partial \phi)}^2$) and the Galileon term
and is shift symmetric for the field absolute value being larger than a particular value.
The spectator field is assumed to be moving in the shift symmetric regime during inflationary era
and to reach the region of shift symmetry breaking in the post-inflationary epoch.
In the shift symmetric region, the energy density of the field behaves as a cosmological constant 
and gradually increases compared to the dominant component.
Thanks to the explicit breaking of the shift symmetry in the Lagrangian, 
this field has a graceful exit from driving a second de Sitter expansion by this field.
After the field passes through the transition regime, it enters another shift symmetric region
where the field behaves as a stiff matter whose equation of state obeys $P_\phi=\rho_\phi$.
By analytically solving the background equations of motion of the spectator field both during and after inflation,
we have shown that contribution of this field to the total energy density becomes maximal  
during the phase of shift symmetry breaking, after which its fraction decays in proportion to $a^{-2}$.

We then evaluated generation and evolution of the field perturbation on de Sitter background.
Adopting the energy scale of inflation recently suggested by the BICEP2 experiment, 
we investigated the parameter space where stochatic effects and the second inflation,
both of which spoil the scenario we consider, are circumvented.
These conditions require that the characteristic energy scale of the spectator field is
larger than the Hubble parameter during inflation.
Having computed the amplitude of the field perturbation,
we then calculated the curvature perturbation sourced by the spectator field
by means of the $\delta N$ formalism. 
We found that a temporal enhancement of the curvature perturbation is realized when the field
is in the transition regime.
The enhancement persists over about one Hubble time.
The amplitude of the enhanced curvature perturbation is much smaller than unity if the Galileon
term determines the background dynamics of the spectator field
during inflation.
In this case, there is no observable that we can probe this scenario.
On the other hand, if the Galileon term is always subdominant, due to reduction of the sound speed
of the spectator field compared to the former case, field perturbation is amplified and
the resultant enhanced curvature perturbation can be as large as unity for some range of parameter space.  
When the perturbation modes during the temporal enhancement reenter the Hubble horizon,
primordial black holes are efficiently produced.
Once produced, they may dominate the late time Universe and after that, they evaporate by the
Hawking radiation and the Universe is again reheated from which all the relevant matter in today's 
Universe is produced.
Dominance of PBHs and the reheating by their evaporation modify the expansion history of the primordial 
Universe, which is left in the shape of the power spectrum of the inflationary tensor mode.
Characteristic feature of the tensor mode spectrum is the red-tilt ($\Omega_{\rm gw} \propto f^{-2}$)
in a frequency range corresponding to PBH dominated epoch and the (almost) flat spectrum outside 
that range.
Interestingly, this range naturally falls into the DECIGO band and thus our scenario can be 
tested by measuring the primordial tensor modes.

Finally, the other interesting observable which we have not considered in this paper is
the gravitational waves from the second order coupling of the spectator field fluctuations. 
In fact, the source term of the second-order gravitational waves can come from both the
adiabatic curvature perturbation and the anisotropic stress caused by the kinetic energy of the curvaton field \cite{Kawasaki:2012wr}.
The former source is independent of the underlying model and has been studied in \cite{Suyama:2011pu}.
Given that large tensor modes generated by spectator fields with
generalized kinetic energy during inflation are found \cite{Biagetti:2013kwa},
a formal investigation of the second-order gravitational waves
generated by a spectator field with higher-order kinetic interactions is definitly desirable
as a future research.
\section*{ACKNOWLEDGMENTS}
The authors would like to thank Sachiko Kuroyanagi for providing us the sensitivity
curves of DECIGO and ultimate-DECIGO.
The authors would also like to thank Tsutomu Kobayashi, Tomohiro Nakama and
Yuki Watanabe for useful comments and discussions. 
Y. P. W. is supported by National Science Council Overseas Project for Post Graduate Research
(NSC-102-2917-I-007-030). 
This work was supported by JSPS Grant-in-Aid for Scientific Research 23340058(JY)
and Grant-in-Aid for Scientific Research on Innovative Areas N0. 25103505(TS).

\bibliographystyle{apsrev}
\bibliography{temporal}

\begin{thebibliography}{44}
\bibitem[{sac()}]{sachikuro}
\bibinfo{howpublished}{We would like to thank Sachiko kuroyanagi for providing
  us the sensitivity curves of DECIGO and ultimate-DECIGO.}
\end{thebibliography}

\end{document}